\documentclass[titlepage,twocolumn,superscriptaddress,prl]{revtex4-1}
\usepackage{amsmath}
\usepackage{amsfonts}
\usepackage{amssymb}
\usepackage{graphicx}
\usepackage{color}
\newcommand{\K}{\mathrm{K}^+}
\newcommand{\Cl}{\mathrm{Cl}^-}
\newcommand{\Km}{\mathrm{K}}
\newcommand{\Clm}{\mathrm{Cl}}
\newcommand{\kt}{k_B T}

\begin{document}

\title{Ionic selectivity and filtration from fragmented dehydration in multilayer graphene nanopores}

\author{Subin Sahu}
\affiliation{Center for Nanoscale Science and Technology, National Institute of Standards and Technology, Gaithersburg, MD 20899}
\affiliation{Maryland Nanocenter, University of Maryland, College Park, MD 20742}
\affiliation{Department of Physics, Oregon State University, Corvallis, OR 97331}

\author{Michael Zwolak}
\email[\textbf{Corresponding Author:} ]{mpz@nist.gov\\}
\affiliation{Center for Nanoscale Science and Technology, National Institute of Standards and Technology, Gaithersburg, MD 20899}

\begin{abstract}
Selective ion transport is a hallmark of biological ion channel behavior but is a major challenge to engineer into artificial membranes. Here, we demonstrate, with all-atom molecular dynamics simulations, that bare graphene nanopores yield measurable ion selectivity that varies over one to two orders of magnitude simply by changing the pore radius and number of graphene layers. Monolayer graphene does not display dehydration-induced selectivity until the pore radius is small enough to exclude the first hydration layer from inside the pore. Bi- and tri-layer graphene, though, display such selectivity already for a pore size that barely encroaches on the first hydration layer, which is due to the more significant water loss from the second hydration layer. Measurement of selectivity and activation barriers from both first and second hydration layer barriers will help elucidate the behavior of biological ion channels. Moreover, the energy barriers responsible for selectivity -- while small on the scale of hydration energies -- are already relatively large, i.e., many $\kt$. For separation of ions from water, therefore, one can exchange longer, larger radius pores for shorter, smaller radius pores, giving a practical method for maintaining exclusion efficiency while enhancing other properties (e.g., water throughput).
\end{abstract}
\maketitle


\begin{figure*}
\includegraphics{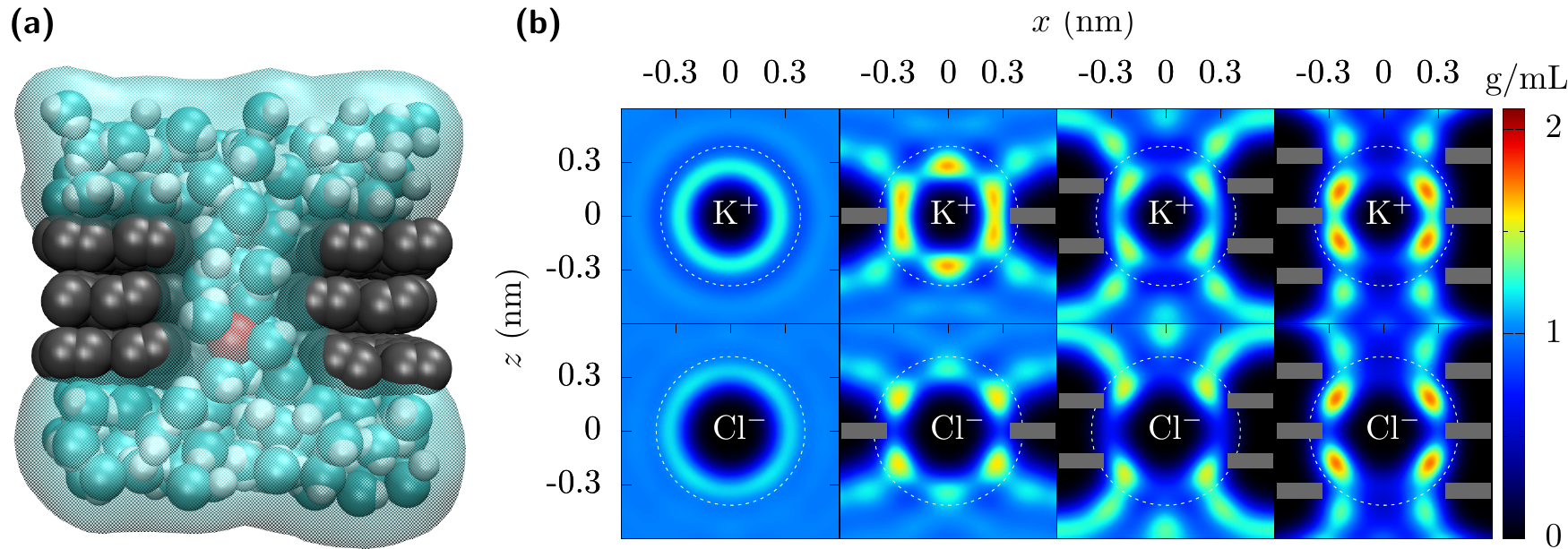}
\caption{{\bf Dehydration of ions going through multilayer graphene pores. }{\bf (a)} A nanopore through trilayer graphene. As $\K$ (red) translocates through the pore, it retains only part of its hydration. In this case, the pore radius is $r_p = 0.34$ nm and the first hydration layer is essentially complete. The second hydration layer, though, is significantly diminished due to the carbon of the graphene (gray) preventing the water molecules (cyan and white) from fluctuating about 0.5 nm away from the ion, except along the pore axis. {\bf (b)} Water density quantified by its oxygen location around $\K$ and $\Cl$ ions fixed in bulk and in the center of mono-, bi-, and tri-layer graphene (shown as grey bars) pores with radius $r_p = 0.34$ nm. The white dotted circles demarcate the first and the second hydration layers. The first hydration layers remain but acquire some additional structure. The second hydration layer is greatly reduced (see Fig.~\ref{fig:Barriers}, and Fig. S2 and Table S2 in the SI). For this pore size, the free energy barrier due to the second layer dehydration significantly contributes to the ion currents and selectivity.\label{fig:Dehydration} The bi- and tri-layer graphene are AB and ABA stacked, respectively, but similar results occur for perfectly aligned multilayer graphene.}
\end{figure*}

\begin{figure}[h!]
\includegraphics{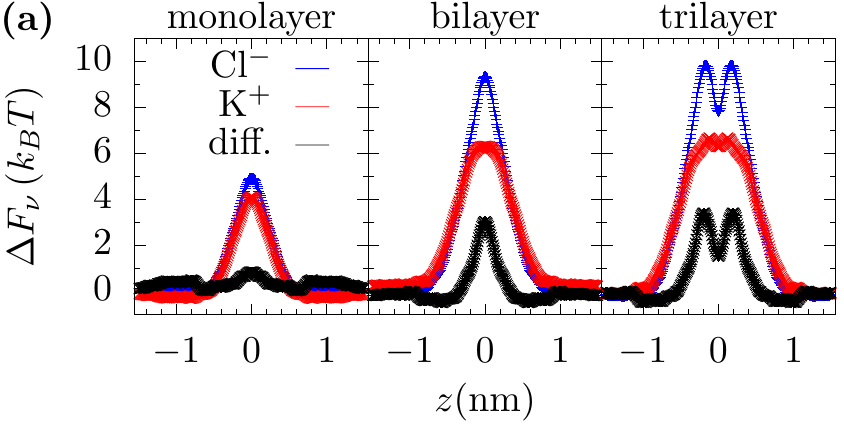}
\includegraphics{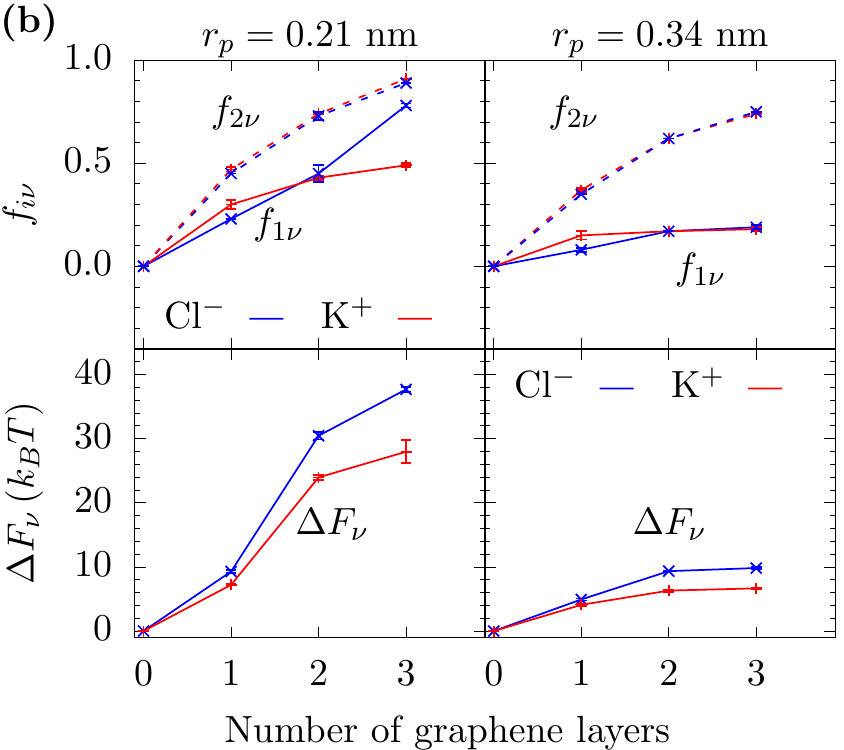}
\caption{{\bf Free energy barriers and dehydration.} {\bf (a)} Free energy barrier versus $\K$ and $\Cl$ location, $z$, on the pore axis as they cross mono-, bi-, and tri-layer graphene pores with radius 0.34 nm. As the number of layers increases, the energy barrier becomes more substantial and a difference between the two ion types appears. {\bf (b)}  Fractional dehydration in the first and second layer ($f_{1\nu}$ and $f_{2\nu}$) for $\K$ and $\Cl$, where the ion is at the position of its free energy maximum in the pore. When the pore radius is less than the first hydration layer radius (about 0.3 nm), then both the first and second hydration layers lose a substantial amount of their water molecules (upper left panel). However, with just a slightly larger pore radius, $r_p=0.34$ nm, the first hydration layer retains most of its water but the second layer still loses a significant number of water molecules (upper right panel). The free energy barriers (lower panels) will increase with the number of graphene layers, as a ``short pore'' interferes less with the hydration than the longer pores. However, while dehydration is the mechanism by which selectivity occurs, water loss is not the sole predictor of selectivity. As Eq.~\ref{eq:fE} shows, one also needs the hydration layer energies. The $\Cl$ ion has a larger hydration energy and, thus, even for the same $f_{i\nu}$, $\Cl$ will be selected against. Error bars are $\pm$1 standard error from five parallel simulations. \label{fig:Barriers} }
\end{figure}

\begin{figure}[h!]
\includegraphics{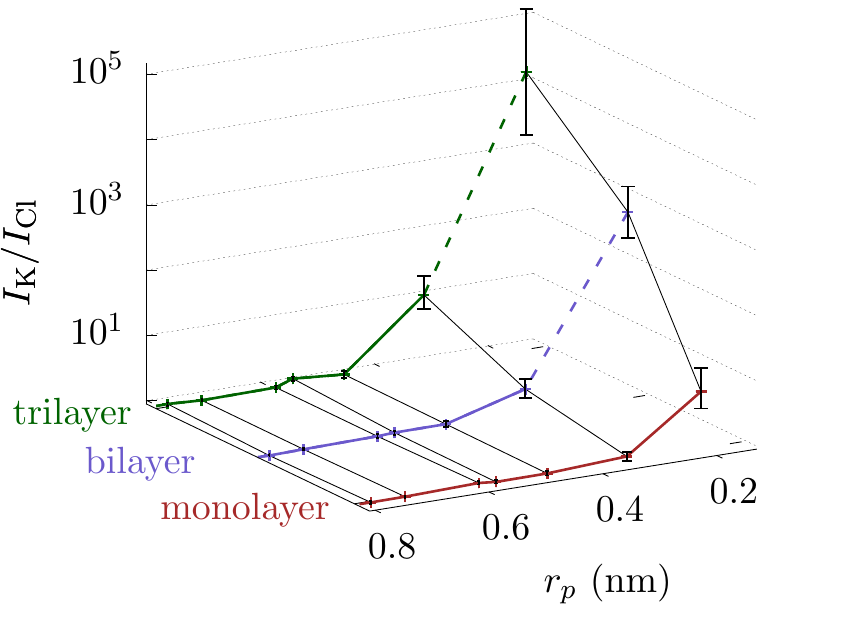}
\caption{{\bf Selectivity of graphene pores.} The selectivity, $I_\Km/I_\Clm$, is at an applied bias of 1 V, although the permeation rates should follow similar trends. Selectivity increases as pore radius decreases and when the number of layers increases. Trilayer graphene with $r_p=0.34$ nm gives a similar selectivity as monolayer graphene with $r_p=0.21$ nm. Moreover, if only ion filtration is of interest, then these two pore sizes can be exchanged. For bi- and tri-layer graphene, we use Eq.~\ref{eq:SelectEst} for $r_p=0.21$ nm, as the currents are too small to reliably determine computationally. Those points have a dashed line connecting them to the remaining plot. The error bars are $\pm$ 1 block standard error (BSE).\label{fig:Selectivity} } 
\end{figure}

Ion transport is vital to physiological processes in the cell~\cite{hille2001,bagal2012,Rasband2010}, where membrane ion channels control ion motion through the interplay of protein structural transitions, precisely placed dipoles and charges, and dehydration. Nanotechnologies seek to mimic and exploit the same physical mechanisms for membrane filtration and desalination. However, biological systems are complex and make use of sophisticated assembly methods, ones that remain difficult to utilize in artificial devices. Recent work, though, on two-dimensional channels in graphene laminates demonstrates ion selectivity~\cite{abraham2017} by constraining the channel height. One-dimensional channels -- pores -- give additional control over the confining geometry, where, for instance, recent theoretical results~\cite{Sahu2017Nano,*Sahu2016} show that experiments on sub-nanoscale, monolayer graphene pores likely display dehydration-only selectivity~\cite{OHern2014}.

Using all-atom molecular dynamics (MD) simulation and theoretical arguments, we show that the most fundamental of all processes -- dehydration of ions -- can be reliably tuned in bare graphene nanopores by controlling only the pore radius and number of graphene layers. This gives rise to selectivity across one to two orders of magnitude before ion currents drop to unmeasurable levels. This range of achievable selectivities is possible due to the ability to separately control the pore radius and length at the nanoscale, i.e., in the regime that influences the hydration layers via the confinement. 
 
 Figure~\ref{fig:Dehydration} shows how the hydration layers change for mono- to trilayer graphene pores. As an ion goes from bulk into the pore, it can not bring its whole hydration layer with it, but rather some of the water molecules are blocked from entering the pore. The shedding of some of the hydration gives a free energy barrier, a simple estimate of which is,
\begin{equation}
\Delta F_\nu = \sum_i f_{i\nu} E_{i\nu} \label{eq:fE},
\end{equation} 
where, $f_{i\nu}$ ($E_{i\nu}$ ) is the fractional dehydration (energy) in the $i^\mathrm{th}$ hydration layer~\cite{Zwolak09-1,Zwolak10-1}. The fractional dehydration depends on the confinement via the pore radius and length (number of graphene layers), as this reduces the volume available for water to hydrate the ion. That is,  
\begin{equation}
f_{i\nu} = \frac{\Delta n_i}{n_i} \approx \frac{\Delta V_i}{V_i}, \label{eq:frac}
\end{equation}
with the total hydration number $n_i$ and volume $V_i$ of the $i^\mathrm{th}$ hydration layer in bulk and the reduction, $\Delta n_i$ and $\Delta V_i$, of those respective quantities in the pore. The quantity $\Delta V_i$ comes from pure geometric arguments -- it is the volume excluded by the presence of graphene carbon atoms -- and the approximation in Eq.~\ref{eq:frac} agrees well with the loss of water molecules computed from MD simulations [shown in Fig.~\ref{fig:Dehydration}(b)]. For narrow pores that split the hydration layer into two hemispherical caps, one can use the surface area available for waters to hydrate the ion, instead of volumes~\cite{Zwolak09-1,Zwolak10-1,Sahu2017Nano}. The Supplementary Information (SI) contains additional details.

For the radius $r_p=0.34$ nm pore in Fig.~\ref{fig:Dehydration}(a), this simple analytic estimate predicts that there should be a small amount of dehydration in the first layer, increasing when going from mono- to bi-/tri-layer graphene. For the multilayer graphene, though, the second hydration layer is significantly reduced. However, due to the much larger hydration energy of the first layer~\cite{Zwolak10-1}, both hydration layers influence the magnitude of the ion currents and thus the selectivity. Moreover, the contribution to the dehydration free energy barrier from hydration layer $i$ will ``level off'' when the length is greater than about twice its radius, i.e., when part of the hydration layer can no longer reside outside of the pore.

This is exactly what is seen from free energy computations using MD. Fig.~\ref{fig:Barriers}(a) shows the free energy barrier for $\K$ and $\Cl$ moving through the pore. Monolayer graphene interferes very little with the hydration for this pore radius. To the extent that this membrane dehydrates the ions, the remaining water molecule can partially compensate for this effect by more strongly orienting their dipole moment with the ion, see the SI. When the number of layers increases, however, the energy barriers change in size and shape.  For both bi- and tri-layer graphene, the dehydration is more substantial and, when accounting for the larger $\Cl$ hydration energy, it starts to differentiate between the two ions. That is, the relative barriers are predominantly influenced by the hydration energies of the different ions. As the confinement increases -- decreasing the pore radius and increasing the pore length -- more water will be lost from the hydration layers, and ions with larger hydration energies will be more effectively filtered by the pore and selected against. Fig.~\ref{fig:Barriers}(b) shows this effect, i.e., how the dehydration and free energy barriers increase with increasing number of graphene layers. 

The free energy barriers are the primary factor in determining permeation rates and ion currents. For instance, the current in the pore is related to the free energy barrier and electric field $E$ according to~\cite{Zwolak10-1}
\begin{equation}
I_\nu  =  e z_\nu \mu_\nu^\mathrm{eff} E A_p n_\nu e^{-\Delta F_\nu / k_B T},
\end{equation}
where, $e$ is the electric charge, $z_\nu$ the ion valency, $\mu_\nu^\mathrm{eff}$ the effective mobility in the pore, $A_p$ is the area of the pore, $n_\nu$ the bulk ion density, $k_B$ is Boltzmann's constant, and $T$ is the temperature. The factors that contribute to selectivity are $\mu_\nu^\mathrm{eff}$ and $\Delta F_\nu$ (and, to some extent, the accessible area for transport is ion dependent as it relates to hydrated ion size. This can be neglected here.). For atomically thin graphene membranes, one expects that the effective mobility is ill-defined. Even still, its contribution to selectivity should be of order 1 (for instance, the ratio of effective mobilities of $\K$ and $\Cl$ goes from about 1 in bulk to about 1.2 in $\alpha$-hemolysin~\cite{bhattacharya2011}). We can thus estimate selectivity as
\begin{equation}
\frac{I_\Km}{I_\Clm} \approx e^{(\Delta F_\Km-\Delta F_\Clm) / k_B T}. \label{eq:SelectEst}
\end{equation}
This is, however, only an estimate: In addition to the effects just discussed, the energy landscape has some ion-dependent spatial structure (which introduces additional factors into the current), and it changes when a bias is applied. For instance, the applied field orients the water dipoles, which can subsequently chaperone ions across the pore~\cite{Sahu2017Nano}. Eq.~\ref{eq:SelectEst}, though, gives the expected scale for selectivity.

Using nonequilibrium MD, we directly compute $I_\Km/I_\Clm$ where possible and use Eq.~\ref{eq:SelectEst} otherwise.  Fig.~\ref{fig:Selectivity} shows the selectivity for pores of radii ranging from 0.21 nm to 0.79 nm in mono-, bi-, and tri-layer graphene. Just as the above theoretical arguments and free energy simulations indicate, the relative current of $\K$ increases compared to $\Cl$ as the pore radius approaches the hydration. The magnitude of this selectivity depends on the pore radius as well as the number of graphene layers. 
We note that the pores are electrically neutral and contain no dipoles. Hence, the selectivity is due to differences in their hydration energies of the ions. All ion types will thus display mutual selectivity. We also note that chemical functionalization of the pore and of the graphene can modify energy barriers, especially when, e.g., the chemical groups are strongly polar or charged under some ionic conditions. When this occurs, the sign of the charge matters, and anions, for instance, may be excluded from the pore. Thus, the selectivity between cations and anions due to a charged pore will be stronger and observable for larger pores, as seen in Ref.~\onlinecite{rollings2016}. However, the effect we discuss will never-the-less be present between cations, where Eqs.~\ref{eq:fE} and~\ref{eq:frac} can estimate the selectivity.

The selectivity that is measurable experimentally will be limited by the minimum resolvable current. The $\Cl$ current is about 5 pA for the 0.34 nm trilayer pore (see Table S1 in the SI). Currents as low as 1 pA are measurable in experiments \cite{balijepalli2014}; thus a several fold change in selectivity should be detectable as the pore size and length vary. This will enable the experimental extraction of dehydration energy barriers (via the temperature dependence of the current) versus the size (length and radius) of this artificial ``selectivity filter''.

Moreover, this provides a method to control selectivity beyond just changing the pore radius so that, e.g., other aspects of the device can be controlled for. 
According to Ref.~\onlinecite{tanugi2016}, the water flow rate only decreases by about 20~\% when going from mono- to bi- layer graphene when the pore size is kept constant, and there is no additional inter-layer spacing.  Increasing the number of layers to increase selectivity (or ion exclusion overall) will not significantly reduce water flow for applications such as desalination. Moreover, for a given selectivity or ion exclusion, one can use a larger pore with more layers, increasing the overall water throughput (as the area available for transport is larger) and membrane stability. 

These results indicate that to achieve a given selectivity, one can exchange a $r_p=0.21$ nm monolayer pore with a trilayer pore of a larger radius ($r_p=0.34$ nm). These pore sizes are both clearly small, but this indicates that, {\em when dealing with nanostructures}, there is flexibility on how to create the desired ion exclusion. Pore sizes are controllable with individual pores fabricated with transmission electron microscopes \cite{garaj2010,Schneider2010,merchant2010} and techniques are under development to fabricate large scale membranes with precise control \cite{OHern2014,Jain2015}. Moreover, we examine only pores with high symmetry. Varying the aspect ratio and the shape of the pore can further tune the conductance and the ion selectivity provided the lateral dimensions of the pore are on the scale of hydration. In any case, layering gives an additional, discrete ``knob'' to tune selectivity and exclusion.

Ion transport through sub-nanometer channels, where dehydration is inevitable, is a key process in biology. Ion transport at this scale is also increasingly important in applications, such as nanopore sequencing (both ionic~\cite{Kasianowicz1996-1,clarke2009,sathe2011} and electronic~\cite{Zwolak05-1,Lagerqvist06-1,Zwolak08-1}), desalination \cite{lee2011} and filtration \cite{karan2015}. Graphene membranes and laminates, as well as other atomically thick membranes, are playing a central role, where selective ion transport and ion exclusion is desired~\cite{OHern2014, rollings2016,Sahu2017Nano, walker2017,Surwade2015,Joshi2014,abraham2017}. Moreover, fundamental studies demonstrate the possibilities of seeing ionic analogs of electric phenomena, such as quantized ionic conductance~\cite{Zwolak09-1,Zwolak10-1} and ionic Coulomb blockade \cite{krems2013,feng2016}. 

Our results form the basis for engineering and understanding selectivity and exclusion with multilayer graphene pores, where both the radial and longitudinal lengths can be controlled at the sub-nanoscale level. This is a feat not easily achievable with other approaches, e.g., solid state~\cite{Zwolak09-1,Zwolak10-1} or carbon nanotubes~\cite{song2009,richards2012,richards2012quantifying} (despite some success in making ultra-thin solid state pores\cite{rodriguez2015}). 
Moreover, examining pores with intermediate pore radii (but ``non-circular'') may show that there is a notion of {\em quantized ionic selectivity}, that for, e.g., trilayer graphene, as the pore radius is reduced, the second hydration layer first gives rise to selectivity, and then the first layer (see the SI for an extended discussion). 
Channel/pore geometry gives a range of possibilities for designing selective pores and experimentally delineating the role of dehydration (to, e.g., understand more complex biological ion channels). Chemical functionalization~\cite{sint2008} and other factors give further possibilities for modifying and engineering selective behavior. 

\smallskip{\noindent \bf Methods}\\
We perform all-atom molecular dynamics (MD) simulations using NAMD2 \cite{phillips2005} with the time step of 2 fs and periodic boundary condition in all directions. The water model is rigid TIP3P \cite{jorgensen1983} from the CHARMM27 force field. Bi- and tri-layer graphene has AB and ABA stacking, respectively. The real-time current comes from applying a 1 V potential across the simulation cell and counting the ion crossing events. The free energies are from equilibrium MD simulations using the adaptive biasing force (ABF) method \cite{Darve2008,Henin2004}. The SI contains additional details regarding methodology. Movie S1, Movie S2, and Movie S3 show a $\K$ ion translocating through mono-, bi-, and tri-layer graphene pores, respectively.

\smallskip{\noindent \bf Acknowledgments}\\
We thank J. Elenewski and M. Di Ventra for helpful comments. S. Sahu acknowledges support under the Cooperative Research Agreement between the University of Maryland and the National Institute of Standards and Technology Center for Nanoscale Science and Technology, Award  70NANB14H209, through the University of Maryland.


\begin{thebibliography}{39}%
\makeatletter
\providecommand \@ifxundefined [1]{%
 \@ifx{#1\undefined}
}%
\providecommand \@ifnum [1]{%
 \ifnum #1\expandafter \@firstoftwo
 \else \expandafter \@secondoftwo
 \fi
}%
\providecommand \@ifx [1]{%
 \ifx #1\expandafter \@firstoftwo
 \else \expandafter \@secondoftwo
 \fi
}%
\providecommand \natexlab [1]{#1}%
\providecommand \enquote  [1]{``#1''}%
\providecommand \bibnamefont  [1]{#1}%
\providecommand \bibfnamefont [1]{#1}%
\providecommand \citenamefont [1]{#1}%
\providecommand \href@noop [0]{\@secondoftwo}%
\providecommand \href [0]{\begingroup \@sanitize@url \@href}%
\providecommand \@href[1]{\@@startlink{#1}\@@href}%
\providecommand \@@href[1]{\endgroup#1\@@endlink}%
\providecommand \@sanitize@url [0]{\catcode `\\12\catcode `\$12\catcode
  `\&12\catcode `\#12\catcode `\^12\catcode `\_12\catcode `\%12\relax}%
\providecommand \@@startlink[1]{}%
\providecommand \@@endlink[0]{}%
\providecommand \url  [0]{\begingroup\@sanitize@url \@url }%
\providecommand \@url [1]{\endgroup\@href {#1}{\urlprefix }}%
\providecommand \urlprefix  [0]{URL }%
\providecommand \Eprint [0]{\href }%
\providecommand \doibase [0]{http://dx.doi.org/}%
\providecommand \selectlanguage [0]{\@gobble}%
\providecommand \bibinfo  [0]{\@secondoftwo}%
\providecommand \bibfield  [0]{\@secondoftwo}%
\providecommand \translation [1]{[#1]}%
\providecommand \BibitemOpen [0]{}%
\providecommand \bibitemStop [0]{}%
\providecommand \bibitemNoStop [0]{.\EOS\space}%
\providecommand \EOS [0]{\spacefactor3000\relax}%
\providecommand \BibitemShut  [1]{\csname bibitem#1\endcsname}%
\let\auto@bib@innerbib\@empty
\bibitem [{\citenamefont {Hille}(2001)}]{hille2001}%
  \BibitemOpen
  \bibfield  {author} {\bibinfo {author} {\bibfnamefont {B.}~\bibnamefont
  {Hille}},\ }\href@noop {} {\emph {\bibinfo {title} {Ion channels of excitable
  membranes}}},\ Vol.\ \bibinfo {volume} {507}\ (\bibinfo  {publisher} {Sinauer
  Sunderland, MA},\ \bibinfo {year} {2001})\BibitemShut {NoStop}%
\bibitem [{\citenamefont {Bagal}\ \emph {et~al.}(2012)\citenamefont {Bagal},
  \citenamefont {Brown}, \citenamefont {Cox}, \citenamefont {Omoto},
  \citenamefont {Owen}, \citenamefont {Pryde}, \citenamefont {Sidders},
  \citenamefont {Skerratt}, \citenamefont {Stevens}, \citenamefont {Storer},\
  and\ \citenamefont {Swain}}]{bagal2012}%
  \BibitemOpen
  \bibfield  {author} {\bibinfo {author} {\bibfnamefont {S.~K.}\ \bibnamefont
  {Bagal}}, \bibinfo {author} {\bibfnamefont {A.~D.}\ \bibnamefont {Brown}},
  \bibinfo {author} {\bibfnamefont {P.~J.}\ \bibnamefont {Cox}}, \bibinfo
  {author} {\bibfnamefont {K.}~\bibnamefont {Omoto}}, \bibinfo {author}
  {\bibfnamefont {R.~M.}\ \bibnamefont {Owen}}, \bibinfo {author}
  {\bibfnamefont {D.~C.}\ \bibnamefont {Pryde}}, \bibinfo {author}
  {\bibfnamefont {B.}~\bibnamefont {Sidders}}, \bibinfo {author} {\bibfnamefont
  {S.~E.}\ \bibnamefont {Skerratt}}, \bibinfo {author} {\bibfnamefont {E.~B.}\
  \bibnamefont {Stevens}}, \bibinfo {author} {\bibfnamefont {R.~I.}\
  \bibnamefont {Storer}}, \ and\ \bibinfo {author} {\bibfnamefont {N.~A.}\
  \bibnamefont {Swain}},\ }\href@noop {} {\bibfield  {journal} {\bibinfo
  {journal} {J. Med. Chem.}\ }\textbf {\bibinfo {volume} {56}},\ \bibinfo
  {pages} {593} (\bibinfo {year} {2012})}\BibitemShut {NoStop}%
\bibitem [{\citenamefont {Rasband}(2010)}]{Rasband2010}%
  \BibitemOpen
  \bibfield  {author} {\bibinfo {author} {\bibfnamefont {M.~N.}\ \bibnamefont
  {Rasband}},\ }\href@noop {} {\bibfield  {journal} {\bibinfo  {journal}
  {Nature Education}\ }\textbf {\bibinfo {volume} {3}},\ \bibinfo {pages} {41}
  (\bibinfo {year} {2010})}\BibitemShut {NoStop}%
\bibitem [{\citenamefont {Abraham}\ \emph {et~al.}(2017)\citenamefont
  {Abraham}, \citenamefont {Vasu}, \citenamefont {Williams}, \citenamefont
  {Gopinadhan}, \citenamefont {Su}, \citenamefont {Cherian}, \citenamefont
  {Dix}, \citenamefont {Prestat}, \citenamefont {Haigh}, \citenamefont
  {Grigorieva}, \citenamefont {Carbone}, \citenamefont {Geim},\ and\
  \citenamefont {Nair}}]{abraham2017}%
  \BibitemOpen
  \bibfield  {author} {\bibinfo {author} {\bibfnamefont {J.}~\bibnamefont
  {Abraham}}, \bibinfo {author} {\bibfnamefont {K.~S.}\ \bibnamefont {Vasu}},
  \bibinfo {author} {\bibfnamefont {C.~D.}\ \bibnamefont {Williams}}, \bibinfo
  {author} {\bibfnamefont {K.}~\bibnamefont {Gopinadhan}}, \bibinfo {author}
  {\bibfnamefont {Y.}~\bibnamefont {Su}}, \bibinfo {author} {\bibfnamefont
  {C.~T.}\ \bibnamefont {Cherian}}, \bibinfo {author} {\bibfnamefont
  {J.}~\bibnamefont {Dix}}, \bibinfo {author} {\bibfnamefont {E.}~\bibnamefont
  {Prestat}}, \bibinfo {author} {\bibfnamefont {S.~J.}\ \bibnamefont {Haigh}},
  \bibinfo {author} {\bibfnamefont {I.~V.}\ \bibnamefont {Grigorieva}},
  \bibinfo {author} {\bibfnamefont {P.}~\bibnamefont {Carbone}}, \bibinfo
  {author} {\bibfnamefont {A.~K.}\ \bibnamefont {Geim}}, \ and\ \bibinfo
  {author} {\bibfnamefont {R.~R.}\ \bibnamefont {Nair}},\ }\href@noop {}
  {\bibfield  {journal} {\bibinfo  {journal} {Nat. Nanotechnol.}\ }\textbf
  {\bibinfo {volume} {12}},\ \bibinfo {pages} {546} (\bibinfo {year}
  {2017})}\BibitemShut {NoStop}%
\bibitem [{\citenamefont {Sahu}\ \emph {et~al.}(2017)\citenamefont {Sahu},
  \citenamefont {Di~Ventra},\ and\ \citenamefont {Zwolak}}]{Sahu2017Nano}%
  \BibitemOpen
  \bibfield  {author} {\bibinfo {author} {\bibfnamefont {S.}~\bibnamefont
  {Sahu}}, \bibinfo {author} {\bibfnamefont {M.}~\bibnamefont {Di~Ventra}}, \
  and\ \bibinfo {author} {\bibfnamefont {M.}~\bibnamefont {Zwolak}},\ }\href
  {\doibase 10.1021/acs.nanolett.7b01399} {\bibfield  {journal} {\bibinfo
  {journal} {Nano Lett.}\ }\textbf {\bibinfo {volume} {17}},\ \bibinfo {pages}
  {4719} (\bibinfo {year} {2017})}\BibitemShut {NoStop}%
\bibitem [{\citenamefont {Sahu}\ \emph {et~al.}(2016)\citenamefont {Sahu},
  \citenamefont {Di~Ventra},\ and\ \citenamefont {Zwolak}}]{Sahu2016}%
  \BibitemOpen
  \bibfield  {author} {\bibinfo {author} {\bibfnamefont {S.}~\bibnamefont
  {Sahu}}, \bibinfo {author} {\bibfnamefont {M.}~\bibnamefont {Di~Ventra}}, \
  and\ \bibinfo {author} {\bibfnamefont {M.}~\bibnamefont {Zwolak}},\
  }\href@noop {} {\bibfield  {journal} {\bibinfo  {journal} {arXiv:1605.03134}\
  } (\bibinfo {year} {2016})}\BibitemShut {NoStop}%
\bibitem [{\citenamefont {O'Hern}\ \emph {et~al.}(2014)\citenamefont {O'Hern},
  \citenamefont {Boutilier}, \citenamefont {Idrobo}, \citenamefont {Song},
  \citenamefont {Kong}, \citenamefont {Laoui}, \citenamefont {Atieh},\ and\
  \citenamefont {Karnik}}]{OHern2014}%
  \BibitemOpen
  \bibfield  {author} {\bibinfo {author} {\bibfnamefont {S.~C.}\ \bibnamefont
  {O'Hern}}, \bibinfo {author} {\bibfnamefont {M.~S.~H.}\ \bibnamefont
  {Boutilier}}, \bibinfo {author} {\bibfnamefont {J.-C.}\ \bibnamefont
  {Idrobo}}, \bibinfo {author} {\bibfnamefont {Y.}~\bibnamefont {Song}},
  \bibinfo {author} {\bibfnamefont {J.}~\bibnamefont {Kong}}, \bibinfo {author}
  {\bibfnamefont {T.}~\bibnamefont {Laoui}}, \bibinfo {author} {\bibfnamefont
  {M.}~\bibnamefont {Atieh}}, \ and\ \bibinfo {author} {\bibfnamefont
  {R.}~\bibnamefont {Karnik}},\ }\href@noop {} {\bibfield  {journal} {\bibinfo
  {journal} {Nano Lett.}\ }\textbf {\bibinfo {volume} {14}},\ \bibinfo {pages}
  {1234} (\bibinfo {year} {2014})}\BibitemShut {NoStop}%
\bibitem [{\citenamefont {Zwolak}\ \emph {et~al.}(2009)\citenamefont {Zwolak},
  \citenamefont {Lagerqvist},\ and\ \citenamefont {Di~Ventra}}]{Zwolak09-1}%
  \BibitemOpen
  \bibfield  {author} {\bibinfo {author} {\bibfnamefont {M.}~\bibnamefont
  {Zwolak}}, \bibinfo {author} {\bibfnamefont {J.}~\bibnamefont {Lagerqvist}},
  \ and\ \bibinfo {author} {\bibfnamefont {M.}~\bibnamefont {Di~Ventra}},\
  }\href@noop {} {\bibfield  {journal} {\bibinfo  {journal} {Phys. Rev. Lett.}\
  }\textbf {\bibinfo {volume} {103}},\ \bibinfo {pages} {128102} (\bibinfo
  {year} {2009})}\BibitemShut {NoStop}%
\bibitem [{\citenamefont {Zwolak}\ \emph {et~al.}(2010)\citenamefont {Zwolak},
  \citenamefont {Wilson},\ and\ \citenamefont {Di~Ventra}}]{Zwolak10-1}%
  \BibitemOpen
  \bibfield  {author} {\bibinfo {author} {\bibfnamefont {M.}~\bibnamefont
  {Zwolak}}, \bibinfo {author} {\bibfnamefont {J.}~\bibnamefont {Wilson}}, \
  and\ \bibinfo {author} {\bibfnamefont {M.}~\bibnamefont {Di~Ventra}},\
  }\href@noop {} {\bibfield  {journal} {\bibinfo  {journal} {J. Phys.: Condens.
  Matter}\ }\textbf {\bibinfo {volume} {22}},\ \bibinfo {pages} {454126}
  (\bibinfo {year} {2010})}\BibitemShut {NoStop}%
\bibitem [{\citenamefont {Bhattacharya}\ \emph {et~al.}(2011)\citenamefont
  {Bhattacharya}, \citenamefont {Muzard}, \citenamefont {Payet}, \citenamefont
  {Math\'{e}}, \citenamefont {Bockelmann}, \citenamefont {Aksimentiev},\ and\
  \citenamefont {Viasnoff}}]{bhattacharya2011}%
  \BibitemOpen
  \bibfield  {author} {\bibinfo {author} {\bibfnamefont {S.}~\bibnamefont
  {Bhattacharya}}, \bibinfo {author} {\bibfnamefont {J.}~\bibnamefont
  {Muzard}}, \bibinfo {author} {\bibfnamefont {L.}~\bibnamefont {Payet}},
  \bibinfo {author} {\bibfnamefont {J.}~\bibnamefont {Math\'{e}}}, \bibinfo
  {author} {\bibfnamefont {U.}~\bibnamefont {Bockelmann}}, \bibinfo {author}
  {\bibfnamefont {A.}~\bibnamefont {Aksimentiev}}, \ and\ \bibinfo {author}
  {\bibfnamefont {V.}~\bibnamefont {Viasnoff}},\ }\href@noop {} {\bibfield
  {journal} {\bibinfo  {journal} {J. Phys. Chem. C}\ }\textbf {\bibinfo
  {volume} {115}},\ \bibinfo {pages} {4255} (\bibinfo {year}
  {2011})}\BibitemShut {NoStop}%
\bibitem [{\citenamefont {Rollings}\ \emph {et~al.}(2016)\citenamefont
  {Rollings}, \citenamefont {Kuan},\ and\ \citenamefont
  {Golovchenko}}]{rollings2016}%
  \BibitemOpen
  \bibfield  {author} {\bibinfo {author} {\bibfnamefont {R.~C.}\ \bibnamefont
  {Rollings}}, \bibinfo {author} {\bibfnamefont {A.~T.}\ \bibnamefont {Kuan}},
  \ and\ \bibinfo {author} {\bibfnamefont {J.~A.}\ \bibnamefont
  {Golovchenko}},\ }\href@noop {} {\bibfield  {journal} {\bibinfo  {journal}
  {Nat. Commun.}\ }\textbf {\bibinfo {volume} {7}},\ \bibinfo {pages} {11408}
  (\bibinfo {year} {2016})}\BibitemShut {NoStop}%
\bibitem [{\citenamefont {Balijepalli}\ \emph {et~al.}(2014)\citenamefont
  {Balijepalli}, \citenamefont {Ettedgui}, \citenamefont {Cornio},
  \citenamefont {Robertson}, \citenamefont {Cheung}, \citenamefont
  {Kasianowicz},\ and\ \citenamefont {Vaz}}]{balijepalli2014}%
  \BibitemOpen
  \bibfield  {author} {\bibinfo {author} {\bibfnamefont {A.}~\bibnamefont
  {Balijepalli}}, \bibinfo {author} {\bibfnamefont {J.}~\bibnamefont
  {Ettedgui}}, \bibinfo {author} {\bibfnamefont {A.~T.}\ \bibnamefont
  {Cornio}}, \bibinfo {author} {\bibfnamefont {J.~W.}\ \bibnamefont
  {Robertson}}, \bibinfo {author} {\bibfnamefont {K.~P.}\ \bibnamefont
  {Cheung}}, \bibinfo {author} {\bibfnamefont {J.~J.}\ \bibnamefont
  {Kasianowicz}}, \ and\ \bibinfo {author} {\bibfnamefont {C.}~\bibnamefont
  {Vaz}},\ }\href@noop {} {\bibfield  {journal} {\bibinfo  {journal} {ACS
  Nano}\ }\textbf {\bibinfo {volume} {8}},\ \bibinfo {pages} {1547} (\bibinfo
  {year} {2014})}\BibitemShut {NoStop}%
\bibitem [{\citenamefont {Cohen-Tanugi}\ \emph {et~al.}(2016)\citenamefont
  {Cohen-Tanugi}, \citenamefont {Lin},\ and\ \citenamefont
  {Grossman}}]{tanugi2016}%
  \BibitemOpen
  \bibfield  {author} {\bibinfo {author} {\bibfnamefont {D.}~\bibnamefont
  {Cohen-Tanugi}}, \bibinfo {author} {\bibfnamefont {L.-C.}\ \bibnamefont
  {Lin}}, \ and\ \bibinfo {author} {\bibfnamefont {J.~C.}\ \bibnamefont
  {Grossman}},\ }\href@noop {} {\bibfield  {journal} {\bibinfo  {journal} {Nano
  Lett.}\ }\textbf {\bibinfo {volume} {16}},\ \bibinfo {pages} {1027} (\bibinfo
  {year} {2016})}\BibitemShut {NoStop}%
\bibitem [{\citenamefont {Garaj}\ \emph {et~al.}(2010)\citenamefont {Garaj},
  \citenamefont {Hubbard}, \citenamefont {Reina}, \citenamefont {Kong},
  \citenamefont {Branton},\ and\ \citenamefont {Golovchenko}}]{garaj2010}%
  \BibitemOpen
  \bibfield  {author} {\bibinfo {author} {\bibfnamefont {S.}~\bibnamefont
  {Garaj}}, \bibinfo {author} {\bibfnamefont {W.}~\bibnamefont {Hubbard}},
  \bibinfo {author} {\bibfnamefont {A.}~\bibnamefont {Reina}}, \bibinfo
  {author} {\bibfnamefont {J.}~\bibnamefont {Kong}}, \bibinfo {author}
  {\bibfnamefont {D.}~\bibnamefont {Branton}}, \ and\ \bibinfo {author}
  {\bibfnamefont {J.}~\bibnamefont {Golovchenko}},\ }\href@noop {} {\bibfield
  {journal} {\bibinfo  {journal} {Nature}\ }\textbf {\bibinfo {volume} {467}},\
  \bibinfo {pages} {190} (\bibinfo {year} {2010})}\BibitemShut {NoStop}%
\bibitem [{\citenamefont {Schneider}\ \emph {et~al.}(2010)\citenamefont
  {Schneider}, \citenamefont {Kowalczyk}, \citenamefont {Calado}, \citenamefont
  {Pandraud}, \citenamefont {Zandbergen}, \citenamefont {Vandersypen},\ and\
  \citenamefont {Dekker}}]{Schneider2010}%
  \BibitemOpen
  \bibfield  {author} {\bibinfo {author} {\bibfnamefont {G.~F.}\ \bibnamefont
  {Schneider}}, \bibinfo {author} {\bibfnamefont {S.~W.}\ \bibnamefont
  {Kowalczyk}}, \bibinfo {author} {\bibfnamefont {V.~E.}\ \bibnamefont
  {Calado}}, \bibinfo {author} {\bibfnamefont {G.}~\bibnamefont {Pandraud}},
  \bibinfo {author} {\bibfnamefont {H.~W.}\ \bibnamefont {Zandbergen}},
  \bibinfo {author} {\bibfnamefont {L.~M.}\ \bibnamefont {Vandersypen}}, \ and\
  \bibinfo {author} {\bibfnamefont {C.}~\bibnamefont {Dekker}},\ }\href@noop {}
  {\bibfield  {journal} {\bibinfo  {journal} {Nano Lett.}\ }\textbf {\bibinfo
  {volume} {10}},\ \bibinfo {pages} {3163} (\bibinfo {year}
  {2010})}\BibitemShut {NoStop}%
\bibitem [{\citenamefont {Merchant}\ \emph {et~al.}(2010)\citenamefont
  {Merchant}, \citenamefont {Healy}, \citenamefont {Wanunu}, \citenamefont
  {Ray}, \citenamefont {Peterman}, \citenamefont {Bartel}, \citenamefont
  {Fischbein}, \citenamefont {Venta}, \citenamefont {Luo}, \citenamefont
  {Johnson},\ and\ \citenamefont {Drndi{\'c}}}]{merchant2010}%
  \BibitemOpen
  \bibfield  {author} {\bibinfo {author} {\bibfnamefont {C.~A.}\ \bibnamefont
  {Merchant}}, \bibinfo {author} {\bibfnamefont {K.}~\bibnamefont {Healy}},
  \bibinfo {author} {\bibfnamefont {M.}~\bibnamefont {Wanunu}}, \bibinfo
  {author} {\bibfnamefont {V.}~\bibnamefont {Ray}}, \bibinfo {author}
  {\bibfnamefont {N.}~\bibnamefont {Peterman}}, \bibinfo {author}
  {\bibfnamefont {J.}~\bibnamefont {Bartel}}, \bibinfo {author} {\bibfnamefont
  {M.~D.}\ \bibnamefont {Fischbein}}, \bibinfo {author} {\bibfnamefont
  {K.}~\bibnamefont {Venta}}, \bibinfo {author} {\bibfnamefont
  {Z.}~\bibnamefont {Luo}}, \bibinfo {author} {\bibfnamefont {A.~C.}\
  \bibnamefont {Johnson}}, \ and\ \bibinfo {author} {\bibfnamefont
  {M.}~\bibnamefont {Drndi{\'c}}},\ }\href@noop {} {\bibfield  {journal}
  {\bibinfo  {journal} {Nano Lett.}\ }\textbf {\bibinfo {volume} {10}},\
  \bibinfo {pages} {2915} (\bibinfo {year} {2010})}\BibitemShut {NoStop}%
\bibitem [{\citenamefont {Jain}\ \emph {et~al.}(2015)\citenamefont {Jain},
  \citenamefont {Rasera}, \citenamefont {Guerrero}, \citenamefont {Boutilier},
  \citenamefont {O'Hern}, \citenamefont {Idrobo},\ and\ \citenamefont
  {Karnik}}]{Jain2015}%
  \BibitemOpen
  \bibfield  {author} {\bibinfo {author} {\bibfnamefont {T.}~\bibnamefont
  {Jain}}, \bibinfo {author} {\bibfnamefont {B.~C.}\ \bibnamefont {Rasera}},
  \bibinfo {author} {\bibfnamefont {R.~J.~S.}\ \bibnamefont {Guerrero}},
  \bibinfo {author} {\bibfnamefont {M.~S.}\ \bibnamefont {Boutilier}}, \bibinfo
  {author} {\bibfnamefont {S.~C.}\ \bibnamefont {O'Hern}}, \bibinfo {author}
  {\bibfnamefont {J.-C.}\ \bibnamefont {Idrobo}}, \ and\ \bibinfo {author}
  {\bibfnamefont {R.}~\bibnamefont {Karnik}},\ }\href@noop {} {\bibfield
  {journal} {\bibinfo  {journal} {Nat. Nanotechnol.}\ }\textbf {\bibinfo
  {volume} {10}},\ \bibinfo {pages} {1053} (\bibinfo {year}
  {2015})}\BibitemShut {NoStop}%
\bibitem [{\citenamefont {Kasianowicz}\ \emph {et~al.}(1996)\citenamefont
  {Kasianowicz}, \citenamefont {Brandin}, \citenamefont {Branton},\ and\
  \citenamefont {Deamer}}]{Kasianowicz1996-1}%
  \BibitemOpen
  \bibfield  {author} {\bibinfo {author} {\bibfnamefont {J.~J.}\ \bibnamefont
  {Kasianowicz}}, \bibinfo {author} {\bibfnamefont {E.}~\bibnamefont
  {Brandin}}, \bibinfo {author} {\bibfnamefont {D.}~\bibnamefont {Branton}}, \
  and\ \bibinfo {author} {\bibfnamefont {D.~W.}\ \bibnamefont {Deamer}},\
  }\href@noop {} {\bibfield  {journal} {\bibinfo  {journal} {Proc. Natl. Acad.
  Sci. U. S. A.}\ }\textbf {\bibinfo {volume} {93}},\ \bibinfo {pages} {13770}
  (\bibinfo {year} {1996})}\BibitemShut {NoStop}%
\bibitem [{\citenamefont {Clarke}\ \emph {et~al.}(2009)\citenamefont {Clarke},
  \citenamefont {Wu}, \citenamefont {Jayasinghe}, \citenamefont {Patel},
  \citenamefont {Reid},\ and\ \citenamefont {Bayley}}]{clarke2009}%
  \BibitemOpen
  \bibfield  {author} {\bibinfo {author} {\bibfnamefont {J.}~\bibnamefont
  {Clarke}}, \bibinfo {author} {\bibfnamefont {H.-C.}\ \bibnamefont {Wu}},
  \bibinfo {author} {\bibfnamefont {L.}~\bibnamefont {Jayasinghe}}, \bibinfo
  {author} {\bibfnamefont {A.}~\bibnamefont {Patel}}, \bibinfo {author}
  {\bibfnamefont {S.}~\bibnamefont {Reid}}, \ and\ \bibinfo {author}
  {\bibfnamefont {H.}~\bibnamefont {Bayley}},\ }\href@noop {} {\bibfield
  {journal} {\bibinfo  {journal} {Nat. Nanotechnol.}\ }\textbf {\bibinfo
  {volume} {4}},\ \bibinfo {pages} {265} (\bibinfo {year} {2009})}\BibitemShut
  {NoStop}%
\bibitem [{\citenamefont {Sathe}\ \emph {et~al.}(2011)\citenamefont {Sathe},
  \citenamefont {Zou}, \citenamefont {Leburton},\ and\ \citenamefont
  {Schulten}}]{sathe2011}%
  \BibitemOpen
  \bibfield  {author} {\bibinfo {author} {\bibfnamefont {C.}~\bibnamefont
  {Sathe}}, \bibinfo {author} {\bibfnamefont {X.}~\bibnamefont {Zou}}, \bibinfo
  {author} {\bibfnamefont {J.-P.}\ \bibnamefont {Leburton}}, \ and\ \bibinfo
  {author} {\bibfnamefont {K.}~\bibnamefont {Schulten}},\ }\href@noop {}
  {\bibfield  {journal} {\bibinfo  {journal} {ACS Nano}\ }\textbf {\bibinfo
  {volume} {5}},\ \bibinfo {pages} {8842} (\bibinfo {year} {2011})}\BibitemShut
  {NoStop}%
\bibitem [{\citenamefont {Zwolak}\ and\ \citenamefont
  {Di~Ventra}(2005)}]{Zwolak05-1}%
  \BibitemOpen
  \bibfield  {author} {\bibinfo {author} {\bibfnamefont {M.}~\bibnamefont
  {Zwolak}}\ and\ \bibinfo {author} {\bibfnamefont {M.}~\bibnamefont
  {Di~Ventra}},\ }\href@noop {} {\bibfield  {journal} {\bibinfo  {journal}
  {Nano Lett.}\ }\textbf {\bibinfo {volume} {5}},\ \bibinfo {pages} {421}
  (\bibinfo {year} {2005})}\BibitemShut {NoStop}%
\bibitem [{\citenamefont {Lagerqvist}\ \emph {et~al.}(2006)\citenamefont
  {Lagerqvist}, \citenamefont {Zwolak},\ and\ \citenamefont
  {DiVentra}}]{Lagerqvist06-1}%
  \BibitemOpen
  \bibfield  {author} {\bibinfo {author} {\bibfnamefont {J.}~\bibnamefont
  {Lagerqvist}}, \bibinfo {author} {\bibfnamefont {M.}~\bibnamefont {Zwolak}},
  \ and\ \bibinfo {author} {\bibfnamefont {M.}~\bibnamefont {DiVentra}},\
  }\href@noop {} {\bibfield  {journal} {\bibinfo  {journal} {Nano Lett.}\
  }\textbf {\bibinfo {volume} {6}},\ \bibinfo {pages} {779} (\bibinfo {year}
  {2006})}\BibitemShut {NoStop}%
\bibitem [{\citenamefont {Zwolak}\ and\ \citenamefont
  {Di~Ventra}(2008)}]{Zwolak08-1}%
  \BibitemOpen
  \bibfield  {author} {\bibinfo {author} {\bibfnamefont {M.}~\bibnamefont
  {Zwolak}}\ and\ \bibinfo {author} {\bibfnamefont {M.}~\bibnamefont
  {Di~Ventra}},\ }\href@noop {} {\bibfield  {journal} {\bibinfo  {journal}
  {Rev. Mod. Phys.}\ }\textbf {\bibinfo {volume} {80}},\ \bibinfo {pages} {141}
  (\bibinfo {year} {2008})}\BibitemShut {NoStop}%
\bibitem [{\citenamefont {Lee}\ \emph {et~al.}(2011)\citenamefont {Lee},
  \citenamefont {Arnot},\ and\ \citenamefont {Mattia}}]{lee2011}%
  \BibitemOpen
  \bibfield  {author} {\bibinfo {author} {\bibfnamefont {K.~P.}\ \bibnamefont
  {Lee}}, \bibinfo {author} {\bibfnamefont {T.~C.}\ \bibnamefont {Arnot}}, \
  and\ \bibinfo {author} {\bibfnamefont {D.}~\bibnamefont {Mattia}},\
  }\href@noop {} {\bibfield  {journal} {\bibinfo  {journal} {J. Membr. Sci.}\
  }\textbf {\bibinfo {volume} {370}},\ \bibinfo {pages} {1} (\bibinfo {year}
  {2011})}\BibitemShut {NoStop}%
\bibitem [{\citenamefont {Karan}\ \emph {et~al.}(2015)\citenamefont {Karan},
  \citenamefont {Jiang},\ and\ \citenamefont {Livingston}}]{karan2015}%
  \BibitemOpen
  \bibfield  {author} {\bibinfo {author} {\bibfnamefont {S.}~\bibnamefont
  {Karan}}, \bibinfo {author} {\bibfnamefont {Z.}~\bibnamefont {Jiang}}, \ and\
  \bibinfo {author} {\bibfnamefont {A.~G.}\ \bibnamefont {Livingston}},\
  }\href@noop {} {\bibfield  {journal} {\bibinfo  {journal} {Science}\ }\textbf
  {\bibinfo {volume} {348}},\ \bibinfo {pages} {1347} (\bibinfo {year}
  {2015})}\BibitemShut {NoStop}%
\bibitem [{\citenamefont {Walker}\ \emph {et~al.}(2017)\citenamefont {Walker},
  \citenamefont {Ubych}, \citenamefont {Saraswat}, \citenamefont {Chalklen},
  \citenamefont {Braeuninger-Weimer}, \citenamefont {Caneva}, \citenamefont
  {Weatherup}, \citenamefont {Hofmann},\ and\ \citenamefont
  {Keyser}}]{walker2017}%
  \BibitemOpen
  \bibfield  {author} {\bibinfo {author} {\bibfnamefont {M.~I.}\ \bibnamefont
  {Walker}}, \bibinfo {author} {\bibfnamefont {K.}~\bibnamefont {Ubych}},
  \bibinfo {author} {\bibfnamefont {V.}~\bibnamefont {Saraswat}}, \bibinfo
  {author} {\bibfnamefont {E.~A.}\ \bibnamefont {Chalklen}}, \bibinfo {author}
  {\bibfnamefont {P.}~\bibnamefont {Braeuninger-Weimer}}, \bibinfo {author}
  {\bibfnamefont {S.}~\bibnamefont {Caneva}}, \bibinfo {author} {\bibfnamefont
  {R.~S.}\ \bibnamefont {Weatherup}}, \bibinfo {author} {\bibfnamefont
  {S.}~\bibnamefont {Hofmann}}, \ and\ \bibinfo {author} {\bibfnamefont
  {U.~F.}\ \bibnamefont {Keyser}},\ }\href@noop {} {\bibfield  {journal}
  {\bibinfo  {journal} {ACS Nano}\ }\textbf {\bibinfo {volume} {11}},\ \bibinfo
  {pages} {1340} (\bibinfo {year} {2017})}\BibitemShut {NoStop}%
\bibitem [{\citenamefont {Surwade}\ \emph {et~al.}(2015)\citenamefont
  {Surwade}, \citenamefont {Smirnov}, \citenamefont {Vlassiouk}, \citenamefont
  {Unocic}, \citenamefont {Veith}, \citenamefont {Dai},\ and\ \citenamefont
  {Mahurin}}]{Surwade2015}%
  \BibitemOpen
  \bibfield  {author} {\bibinfo {author} {\bibfnamefont {S.~P.}\ \bibnamefont
  {Surwade}}, \bibinfo {author} {\bibfnamefont {S.~N.}\ \bibnamefont
  {Smirnov}}, \bibinfo {author} {\bibfnamefont {I.~V.}\ \bibnamefont
  {Vlassiouk}}, \bibinfo {author} {\bibfnamefont {R.~R.}\ \bibnamefont
  {Unocic}}, \bibinfo {author} {\bibfnamefont {G.~M.}\ \bibnamefont {Veith}},
  \bibinfo {author} {\bibfnamefont {S.}~\bibnamefont {Dai}}, \ and\ \bibinfo
  {author} {\bibfnamefont {S.~M.}\ \bibnamefont {Mahurin}},\ }\href@noop {}
  {\bibfield  {journal} {\bibinfo  {journal} {Nat. Nanotechnol.}\ }\textbf
  {\bibinfo {volume} {10}},\ \bibinfo {pages} {459} (\bibinfo {year}
  {2015})}\BibitemShut {NoStop}%
\bibitem [{\citenamefont {Joshi}\ \emph {et~al.}(2014)\citenamefont {Joshi},
  \citenamefont {Carbone}, \citenamefont {Wang}, \citenamefont {Kravets},
  \citenamefont {Su}, \citenamefont {Grigorieva}, \citenamefont {Wu},
  \citenamefont {Geim},\ and\ \citenamefont {Nair}}]{Joshi2014}%
  \BibitemOpen
  \bibfield  {author} {\bibinfo {author} {\bibfnamefont {R.}~\bibnamefont
  {Joshi}}, \bibinfo {author} {\bibfnamefont {P.}~\bibnamefont {Carbone}},
  \bibinfo {author} {\bibfnamefont {F.}~\bibnamefont {Wang}}, \bibinfo {author}
  {\bibfnamefont {V.}~\bibnamefont {Kravets}}, \bibinfo {author} {\bibfnamefont
  {Y.}~\bibnamefont {Su}}, \bibinfo {author} {\bibfnamefont {I.}~\bibnamefont
  {Grigorieva}}, \bibinfo {author} {\bibfnamefont {H.}~\bibnamefont {Wu}},
  \bibinfo {author} {\bibfnamefont {A.}~\bibnamefont {Geim}}, \ and\ \bibinfo
  {author} {\bibfnamefont {R.}~\bibnamefont {Nair}},\ }\href@noop {} {\bibfield
   {journal} {\bibinfo  {journal} {Science}\ }\textbf {\bibinfo {volume}
  {343}},\ \bibinfo {pages} {752} (\bibinfo {year} {2014})}\BibitemShut
  {NoStop}%
\bibitem [{\citenamefont {Krems}\ and\ \citenamefont
  {Di~Ventra}(2013)}]{krems2013}%
  \BibitemOpen
  \bibfield  {author} {\bibinfo {author} {\bibfnamefont {M.}~\bibnamefont
  {Krems}}\ and\ \bibinfo {author} {\bibfnamefont {M.}~\bibnamefont
  {Di~Ventra}},\ }\href@noop {} {\bibfield  {journal} {\bibinfo  {journal} {J.
  Phys.: Condens. Matter}\ }\textbf {\bibinfo {volume} {25}},\ \bibinfo {pages}
  {065101} (\bibinfo {year} {2013})}\BibitemShut {NoStop}%
\bibitem [{\citenamefont {Feng}\ \emph {et~al.}(2016)\citenamefont {Feng},
  \citenamefont {Liu}, \citenamefont {Graf}, \citenamefont {Dumcenco},
  \citenamefont {Kis}, \citenamefont {Di~Ventra},\ and\ \citenamefont
  {Radenovic}}]{feng2016}%
  \BibitemOpen
  \bibfield  {author} {\bibinfo {author} {\bibfnamefont {J.}~\bibnamefont
  {Feng}}, \bibinfo {author} {\bibfnamefont {K.}~\bibnamefont {Liu}}, \bibinfo
  {author} {\bibfnamefont {M.}~\bibnamefont {Graf}}, \bibinfo {author}
  {\bibfnamefont {D.}~\bibnamefont {Dumcenco}}, \bibinfo {author}
  {\bibfnamefont {A.}~\bibnamefont {Kis}}, \bibinfo {author} {\bibfnamefont
  {M.}~\bibnamefont {Di~Ventra}}, \ and\ \bibinfo {author} {\bibfnamefont
  {A.}~\bibnamefont {Radenovic}},\ }\href@noop {} {\bibfield  {journal}
  {\bibinfo  {journal} {Nat. Mater.}\ }\textbf {\bibinfo {volume} {15}},\
  \bibinfo {pages} {850} (\bibinfo {year} {2016})}\BibitemShut {NoStop}%
\bibitem [{\citenamefont {Song}\ and\ \citenamefont {Corry}(2009)}]{song2009}%
  \BibitemOpen
  \bibfield  {author} {\bibinfo {author} {\bibfnamefont {C.}~\bibnamefont
  {Song}}\ and\ \bibinfo {author} {\bibfnamefont {B.}~\bibnamefont {Corry}},\
  }\href@noop {} {\bibfield  {journal} {\bibinfo  {journal} {J. Phys. Chem. B}\
  }\textbf {\bibinfo {volume} {113}},\ \bibinfo {pages} {7642} (\bibinfo {year}
  {2009})}\BibitemShut {NoStop}%
\bibitem [{\citenamefont {Richards}\ \emph
  {et~al.}(2012{\natexlab{a}})\citenamefont {Richards}, \citenamefont
  {Sch{\"a}fer}, \citenamefont {Richards},\ and\ \citenamefont
  {Corry}}]{richards2012}%
  \BibitemOpen
  \bibfield  {author} {\bibinfo {author} {\bibfnamefont {L.~A.}\ \bibnamefont
  {Richards}}, \bibinfo {author} {\bibfnamefont {A.~I.}\ \bibnamefont
  {Sch{\"a}fer}}, \bibinfo {author} {\bibfnamefont {B.~S.}\ \bibnamefont
  {Richards}}, \ and\ \bibinfo {author} {\bibfnamefont {B.}~\bibnamefont
  {Corry}},\ }\href@noop {} {\bibfield  {journal} {\bibinfo  {journal} {Small}\
  }\textbf {\bibinfo {volume} {8}},\ \bibinfo {pages} {1701} (\bibinfo {year}
  {2012}{\natexlab{a}})}\BibitemShut {NoStop}%
\bibitem [{\citenamefont {Richards}\ \emph
  {et~al.}(2012{\natexlab{b}})\citenamefont {Richards}, \citenamefont
  {Sch{\"a}fer}, \citenamefont {Richards},\ and\ \citenamefont
  {Corry}}]{richards2012quantifying}%
  \BibitemOpen
  \bibfield  {author} {\bibinfo {author} {\bibfnamefont {L.~A.}\ \bibnamefont
  {Richards}}, \bibinfo {author} {\bibfnamefont {A.~I.}\ \bibnamefont
  {Sch{\"a}fer}}, \bibinfo {author} {\bibfnamefont {B.~S.}\ \bibnamefont
  {Richards}}, \ and\ \bibinfo {author} {\bibfnamefont {B.}~\bibnamefont
  {Corry}},\ }\href@noop {} {\bibfield  {journal} {\bibinfo  {journal} {Phys.
  Chem. Chem. Phys.}\ }\textbf {\bibinfo {volume} {14}},\ \bibinfo {pages}
  {11633} (\bibinfo {year} {2012}{\natexlab{b}})}\BibitemShut {NoStop}%
\bibitem [{\citenamefont {Rodr{\'i}guez-Manzo}\ \emph
  {et~al.}(2015)\citenamefont {Rodr{\'i}guez-Manzo}, \citenamefont {Puster},
  \citenamefont {Nicola{\"i}}, \citenamefont {Meunier},\ and\ \citenamefont
  {Drndi{\'c}}}]{rodriguez2015}%
  \BibitemOpen
  \bibfield  {author} {\bibinfo {author} {\bibfnamefont {J.~A.}\ \bibnamefont
  {Rodr{\'i}guez-Manzo}}, \bibinfo {author} {\bibfnamefont {M.}~\bibnamefont
  {Puster}}, \bibinfo {author} {\bibfnamefont {A.}~\bibnamefont {Nicola{\"i}}},
  \bibinfo {author} {\bibfnamefont {V.}~\bibnamefont {Meunier}}, \ and\
  \bibinfo {author} {\bibfnamefont {M.}~\bibnamefont {Drndi{\'c}}},\
  }\href@noop {} {\bibfield  {journal} {\bibinfo  {journal} {ACS Nano}\
  }\textbf {\bibinfo {volume} {9}},\ \bibinfo {pages} {6555} (\bibinfo {year}
  {2015})}\BibitemShut {NoStop}%
\bibitem [{\citenamefont {Sint}\ \emph {et~al.}(2008)\citenamefont {Sint},
  \citenamefont {Wang},\ and\ \citenamefont {Kr{\'a}l}}]{sint2008}%
  \BibitemOpen
  \bibfield  {author} {\bibinfo {author} {\bibfnamefont {K.}~\bibnamefont
  {Sint}}, \bibinfo {author} {\bibfnamefont {B.}~\bibnamefont {Wang}}, \ and\
  \bibinfo {author} {\bibfnamefont {P.}~\bibnamefont {Kr{\'a}l}},\ }\href@noop
  {} {\bibfield  {journal} {\bibinfo  {journal} {J. Am. Chem. Soc.}\ }\textbf
  {\bibinfo {volume} {130}},\ \bibinfo {pages} {16448} (\bibinfo {year}
  {2008})}\BibitemShut {NoStop}%
\bibitem [{\citenamefont {Phillips}\ \emph {et~al.}(2005)\citenamefont
  {Phillips}, \citenamefont {Braun}, \citenamefont {Wang}, \citenamefont
  {Gumbart}, \citenamefont {Tajkhorshid}, \citenamefont {Villa}, \citenamefont
  {Chipot}, \citenamefont {Skeel}, \citenamefont {Kale},\ and\ \citenamefont
  {Schulten}}]{phillips2005}%
  \BibitemOpen
  \bibfield  {author} {\bibinfo {author} {\bibfnamefont {J.~C.}\ \bibnamefont
  {Phillips}}, \bibinfo {author} {\bibfnamefont {R.}~\bibnamefont {Braun}},
  \bibinfo {author} {\bibfnamefont {W.}~\bibnamefont {Wang}}, \bibinfo {author}
  {\bibfnamefont {J.}~\bibnamefont {Gumbart}}, \bibinfo {author} {\bibfnamefont
  {E.}~\bibnamefont {Tajkhorshid}}, \bibinfo {author} {\bibfnamefont
  {E.}~\bibnamefont {Villa}}, \bibinfo {author} {\bibfnamefont
  {C.}~\bibnamefont {Chipot}}, \bibinfo {author} {\bibfnamefont {R.~D.}\
  \bibnamefont {Skeel}}, \bibinfo {author} {\bibfnamefont {L.}~\bibnamefont
  {Kale}}, \ and\ \bibinfo {author} {\bibfnamefont {K.}~\bibnamefont
  {Schulten}},\ }\href@noop {} {\bibfield  {journal} {\bibinfo  {journal} {J.
  Comput. Chem.}\ }\textbf {\bibinfo {volume} {26}},\ \bibinfo {pages} {1781}
  (\bibinfo {year} {2005})}\BibitemShut {NoStop}%
\bibitem [{\citenamefont {Jorgensen}\ \emph {et~al.}(1983)\citenamefont
  {Jorgensen}, \citenamefont {Chandrasekhar}, \citenamefont {Madura},
  \citenamefont {Impey},\ and\ \citenamefont {Klein}}]{jorgensen1983}%
  \BibitemOpen
  \bibfield  {author} {\bibinfo {author} {\bibfnamefont {W.~L.}\ \bibnamefont
  {Jorgensen}}, \bibinfo {author} {\bibfnamefont {J.}~\bibnamefont
  {Chandrasekhar}}, \bibinfo {author} {\bibfnamefont {J.~D.}\ \bibnamefont
  {Madura}}, \bibinfo {author} {\bibfnamefont {R.~W.}\ \bibnamefont {Impey}}, \
  and\ \bibinfo {author} {\bibfnamefont {M.~L.}\ \bibnamefont {Klein}},\
  }\href@noop {} {\bibfield  {journal} {\bibinfo  {journal} {J. Chem. Phys.}\
  }\textbf {\bibinfo {volume} {79}},\ \bibinfo {pages} {926} (\bibinfo {year}
  {1983})}\BibitemShut {NoStop}%
\bibitem [{\citenamefont {Darve}\ \emph {et~al.}(2008)\citenamefont {Darve},
  \citenamefont {Rodr{\'\i}guez-G{\'o}mez},\ and\ \citenamefont
  {Pohorille}}]{Darve2008}%
  \BibitemOpen
  \bibfield  {author} {\bibinfo {author} {\bibfnamefont {E.}~\bibnamefont
  {Darve}}, \bibinfo {author} {\bibfnamefont {D.}~\bibnamefont
  {Rodr{\'\i}guez-G{\'o}mez}}, \ and\ \bibinfo {author} {\bibfnamefont
  {A.}~\bibnamefont {Pohorille}},\ }\href@noop {} {\bibfield  {journal}
  {\bibinfo  {journal} {J. Chem. Phys.}\ }\textbf {\bibinfo {volume} {128}},\
  \bibinfo {pages} {144120} (\bibinfo {year} {2008})}\BibitemShut {NoStop}%
\bibitem [{\citenamefont {H{\'e}nin}\ and\ \citenamefont
  {Chipot}(2004)}]{Henin2004}%
  \BibitemOpen
  \bibfield  {author} {\bibinfo {author} {\bibfnamefont {J.}~\bibnamefont
  {H{\'e}nin}}\ and\ \bibinfo {author} {\bibfnamefont {C.}~\bibnamefont
  {Chipot}},\ }\href@noop {} {\bibfield  {journal} {\bibinfo  {journal} {J.
  Chem. Phys.}\ }\textbf {\bibinfo {volume} {121}},\ \bibinfo {pages} {2904}
  (\bibinfo {year} {2004})}\BibitemShut {NoStop}%
\end{thebibliography}
%

\end{document}


\title{Ionic selectivity and filtration from fragmented dehydration in multilayer graphene nanopores -- Supplementary Information}

\author{Subin Sahu}
\affiliation{Center for Nanoscale Science and Technology, National Institute of Standards and Technology, Gaithersburg, MD 20899}
\affiliation{Maryland Nanocenter, University of Maryland, College Park, MD 20742}
\affiliation{Department of Physics, Oregon State University, Corvallis, OR 97331}

\author{Michael Zwolak}
\email{mpz@nist.gov}
\affiliation{Center for Nanoscale Science and Technology, National Institute of Standards and Technology, Gaithersburg, MD 20899}

\maketitle
\tableofcontents
\clearpage

\section{Methods}

\subsection{All-atom MD simulations}
We use multilayer graphene with AB stacking with a C-C bond length of $\approx 0.14$ nm and inter-layer distance of $\approx 0.335$ nm. We open a pore of nominal radius $r_n$ at the center of each membrane by removing carbon atoms whose coordinates satisfy the condition  $(x-x_c)^2+(y-y_c)^2<r_n^2$, where ($x_c,\,y_c$) is the center of mass in each graphene membrane. However, the pore radius, $r_p$, is measured from the inner edge of carbon atoms (taken as their van der Waals radius) around the pore. The graphene membrane has a square cross-section of 7.2 nm by 7.2 nm, which we immerse in an aqueous KCl solution of concentration 1 mol/L that extends 5 nm on both sides of the membrane.

We perform all-atom molecular dynamics (MD) simulations using NAMD2 \cite{phillips2005} with a time step of 2 fs and employ periodic boundary condition in all directions. The water model in our simulation is rigid TIP3P \cite{jorgensen1983} from the CHARMM27 force field (previously, we used flexible TIP3P~\cite{sahu2017Nano,*Sahu2016}, which gives similar results but the rigid model allows for more efficient simulations). Non-bonded interactions (van der Waals and electrostatic) have a cutoff of 1.2 nm, but we perform a full electrostatics calculation every 8 fs using particle-mesh Ewald (PME) method~\cite{darden1993}. We prepare the system using VMD \cite{Humphrey1996} and then equilibrate the system using NAMD2. The equilibration steps are (1) minimizing the energy of the system for 4000 steps, (2) heating it to 295 K in another 8 ps, (3) a 1 ns NPT (constant number of particles, pressure and temperature) equilibration using the Nose-Hoover Langevin piston method~\cite{Martyna1994} to raise the pressure to 101325 Pa (i.e., 1 atm), and (4) a 3 ns of NVT (constant number of particles, volume and temperature) equilibration. 

We use real-time, all-atom molecular dynamics simulations to calculate the ionic current through the equilibrated system by applying an electric field perpendicular to the plane of the membrane. We set the Langevin damping rate to 0.2 ps$^{-1}$ for carbon and water (via its oxygen atoms) during these runs. We freeze the carbon atoms at the outer edge of the graphene membrane, but the rest of the carbon atoms in the graphene membrane are only confined by C-C bonds. We averaged the current for a total time of 50 ns to 150 ns depending on the pore size and number of layers.

\begin{figure}[h]
\includegraphics{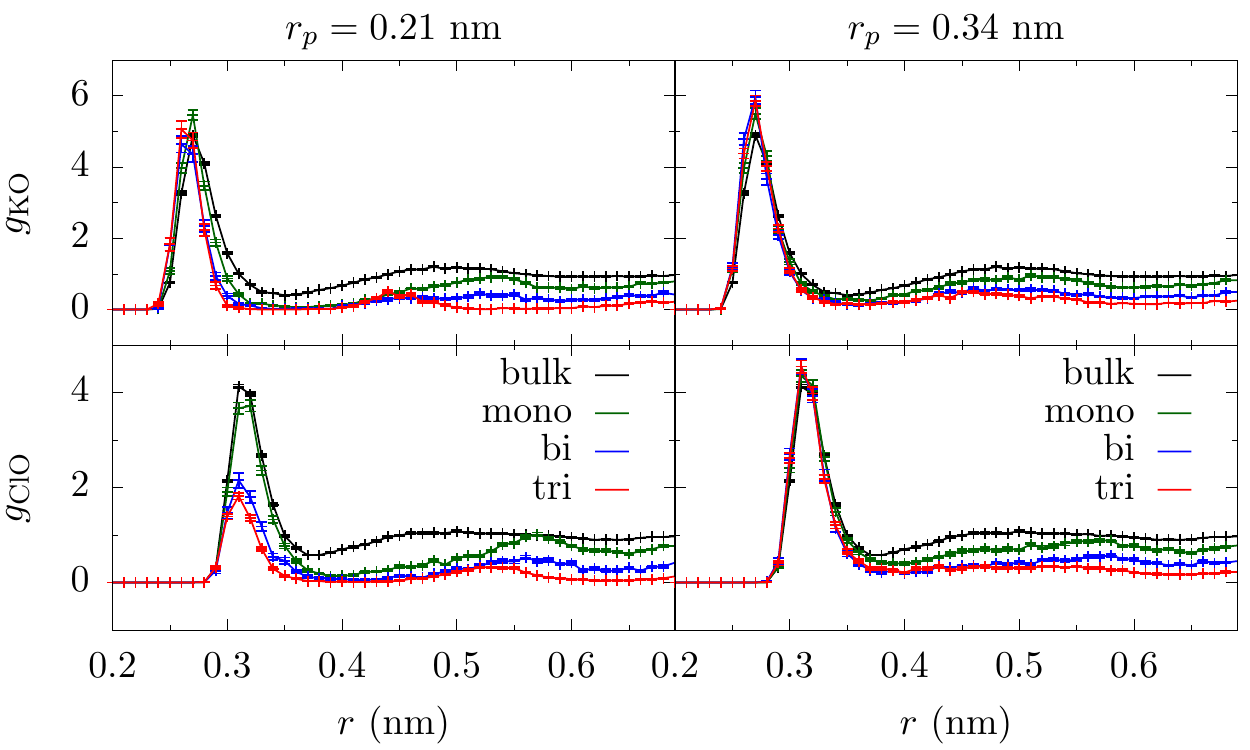}
\caption{\label{rdf}  Radial distribution functions $g_\mathrm{KO}$ and $g_\mathrm{ClO}$ with $\K$ and $\Cl$ ions in bulk and inside the mono-, bi-, and tri-layer graphene pores with radius $r_p = 0.21$ nm and $r_p=0.34$ nm. The bulk ion concentration is maintained at 1 mol/L in each calculation. There is significant dehydration in both the first and second hydration layers in the $r_p=0.21$ nm pore, whereas in the $r_p=0.34$ nm pore dehydration is significant only in the second hydration layer. The error bars are $\pm$ 1 block standard error (BSE).} 
\end{figure}

\begin{figure}[h]
\includegraphics{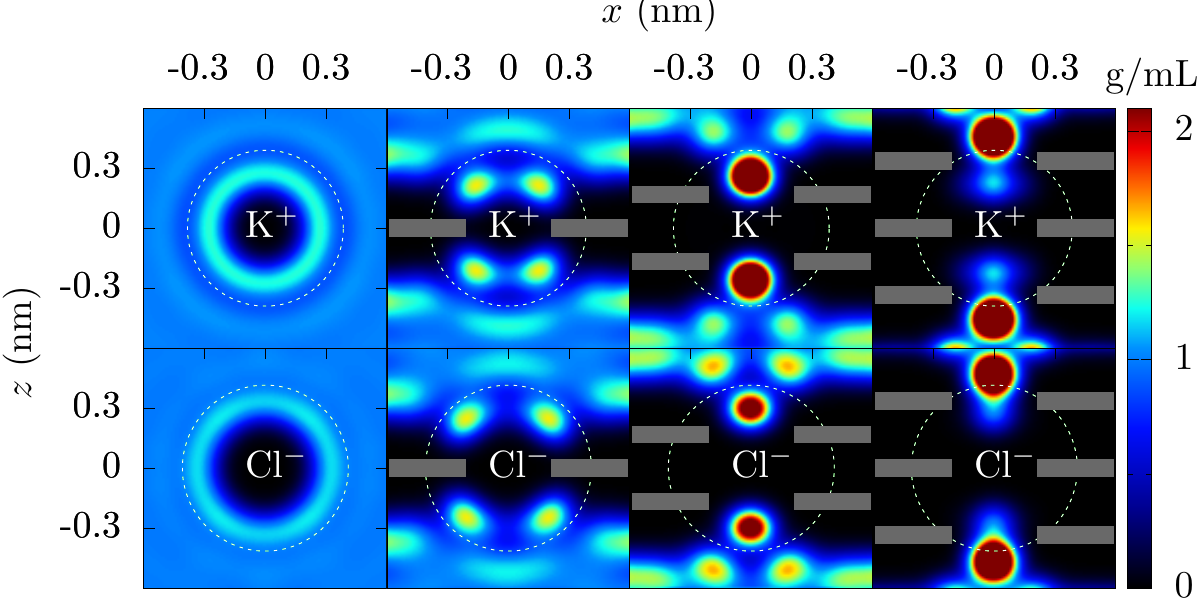}
\caption{\label{solvation}  Water density (within the $y=0$ plane) quantified by its oxygen location around $\K$ and $\Cl$ ions in bulk and mono-, bi-, and tri-layer graphene (shown as gray bars) pores with radius $r_p = 0.21$ nm. The white dotted circles demarcate the first and the second hydration layers. The bi- and tri-layer graphene significantly excludes both the first and second hydration layers. For monolayer graphene, however, most of the hydration layers are still present due to the atomic thickness of the membrane (see Table \ref{WaterTable}). However, the water molecules are more localized than in bulk.} 
\end{figure}

\subsection{Solvation Shells}

To calculate the solvation shells for each ion, we fix the ion in the center of a pore and run equilibrium NVT simulations. Fig.~\ref{rdf} shows the radial distribution functions of oxygen atoms with respect to the ion ($\K$ or $\Cl$) fixed in the bulk and in the center of the pore in mono-, bi-, and tri-layer graphene. Fig.~\ref{solvation} shows the solvation shell around $\K$ and $\Cl$ ions fixed at the center of 0.21 nm pore on mono-, bi-, and tri-layer graphene. A similar plot for 0.34 nm pore is shown in Fig.~1(b) of the main text. These plots show that in monolayer graphene, the ion at the center of the pore can maintain most of its first hydration shell. However, in bi- and tri-layer graphene there is a greater loss of water from first hydration layer. The dehydration is even stronger in the second hydration layer, losing about 50~\%, 80~\%, and 90~\% of water molecules in mono-, bi-, and tri-layer graphene, respectively.  The water molecules around the ion in the pore are spatially localized, thus giving fragmented solvation shells. We note that in Fig.~2(b) of the main text, we calculate the fractional dehydration with the ion within a distance 0.1 nm of its free energy maximum position along $z$-axis, as this is the most relevant location in determining ion transport.

\begin{figure}[h]
\includegraphics{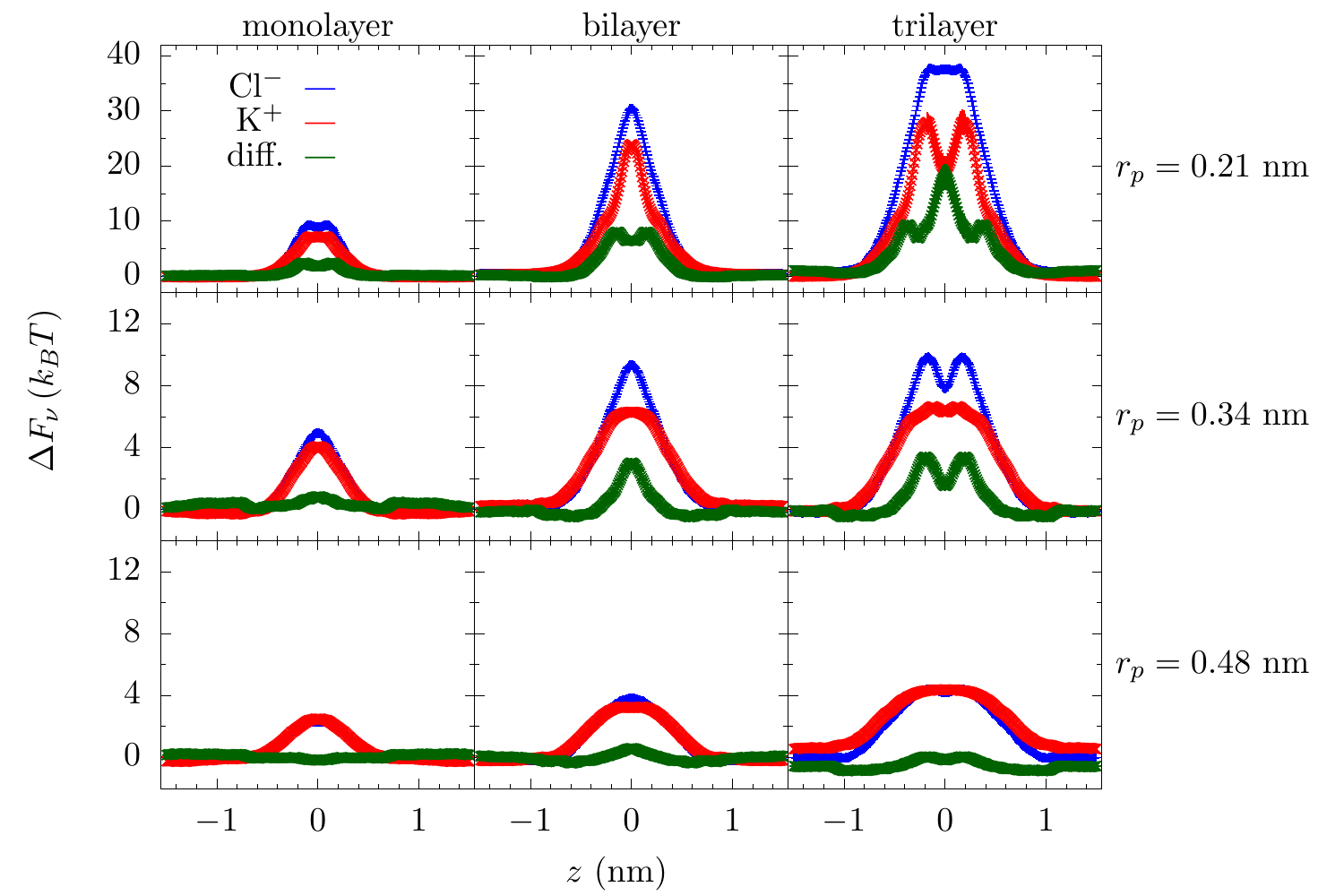}
\caption{\label{pmf} Free energy barrier for $\K$ (red line) ion, $\Cl$ (blue line) ion, and their difference (green line) to translocate through the pore versus the $z$-location for radii 0.21 nm, 0.34 nm, and 0.48 nm pores in mono-, bi-, and tri-layer graphene. The free energy barriers, as well as their difference, increase with decreasing pore radius and with increasing number of graphene layers, thus making the pore more selective. Error bars are $\pm$1 standard error from five parallel simulations. }
\end{figure}

\begin{figure}[h!]
\includegraphics{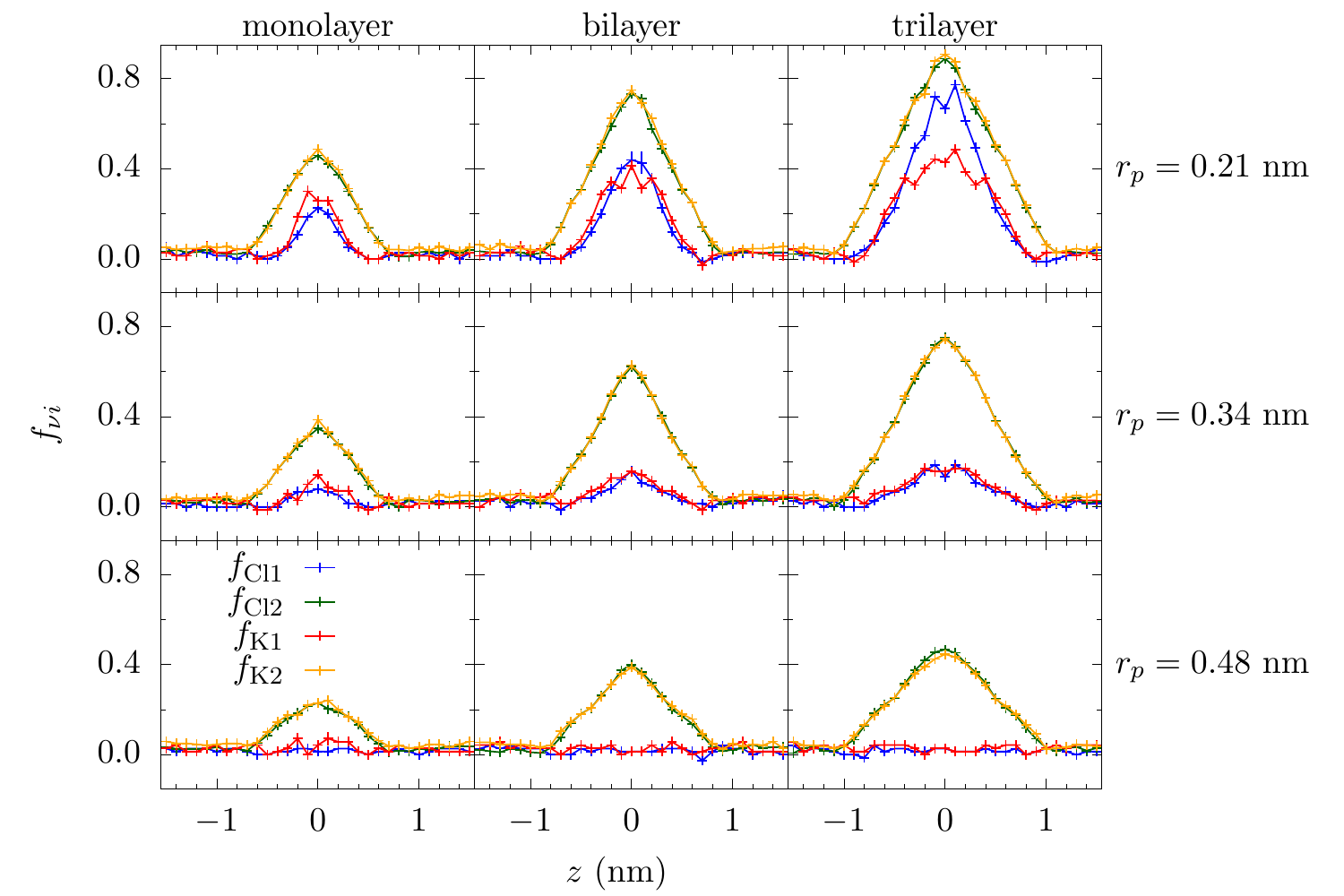}
\caption{\label{hydration} Fractional dehydration in the first and the second hydration layers for $\K$ and $\Cl$ ions translocating through pores of radius 0.21 nm, 0.34 nm, and 0.48 nm in mono-, bi-, and tri-layer graphene. These results come from the same simulation as used to compute the free energy barrier. Just like the free energy barrier, the dehydration increases with the decrease in the pore radius and with the increase in the number of graphene layers. Fractional dehydration is always smaller in the first hydration layer compare to the second hydration layer. However, due to the larger energy of the first hydration layer, it still has a large contribution to the free energy barrier. Error bars are $\pm$1 standard error from five parallel simulations.}
\end{figure}

\subsection{Free Energy Calculations}
We calculate the free energy profile of an ion crossing the pore by using the adaptive biasing force (ABF) method~\cite{Darve2008, Henin2004} as implemented in NAMD2. We compute the free energy barrier within a cylinder of radius $r_p$ and height of 3 nm centered at the origin. Fig.~\ref{pmf} shows the free-energy profile for both $\K$ and $\Cl$ ions and the difference in the free energies of these two ions along the z-axis. The free energy barrier for each ion increases as we decrease the pore radius or increase the number of graphene layers. Also, the difference in the free energy barriers of $\K$ and $\Cl$ increases for decreasing pore radius and increasing number of graphene layers. The free energy barriers appear due to dehydration of ions in the pore (see Fig.~\ref{hydration}). As pore radius decreases and the number of graphene layer increases, the fractional dehydration in the solvation shell of ion increases, as shown in Fig.~\ref{hydration} and Fig.~2(b) of the main text.

\begin{figure}
\includegraphics{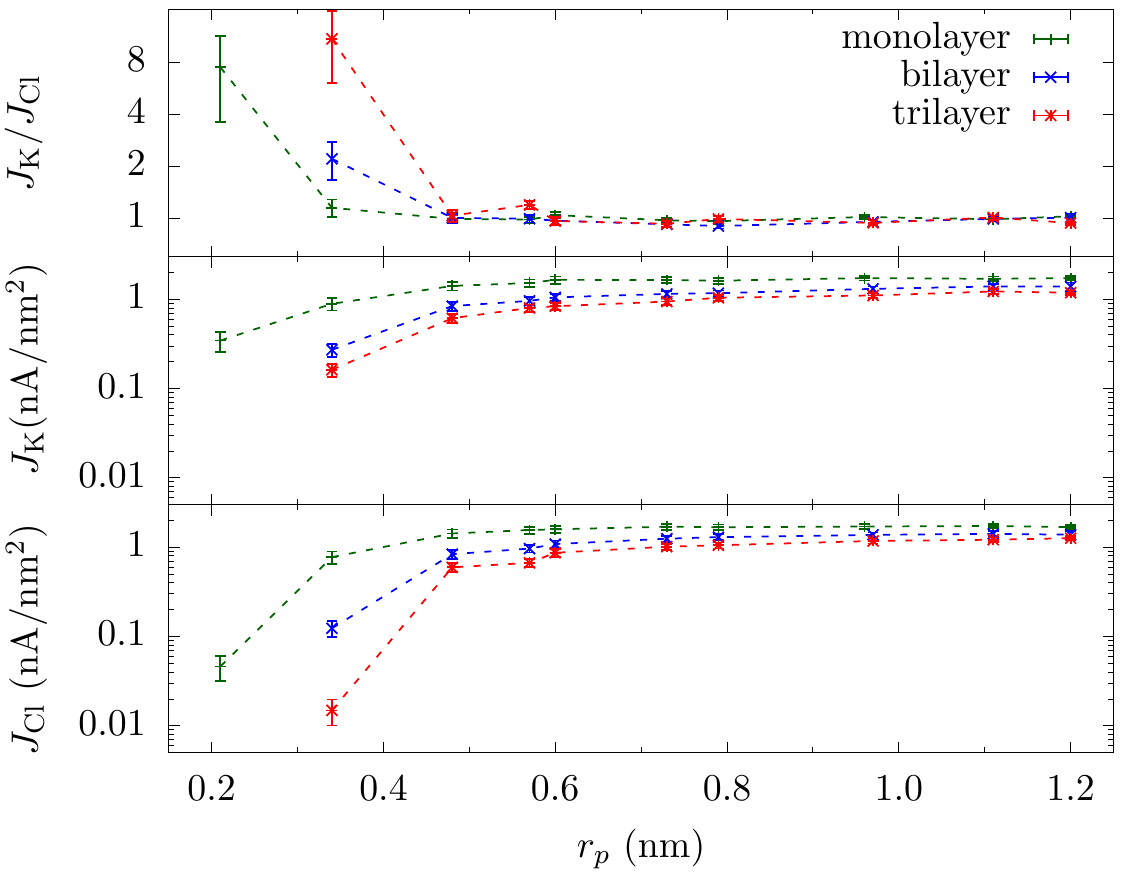}
\caption{\label{current} Current densities in the pore, $J_\Km=I_\Km/\pi r_p^2$ and $J_\Clm=I_\Clm/\pi r_p^2$, and their ratio versus $r_p$ for mono-, bi-, and tri-layer graphene membranes. There is no selectivity ($I_\Km/I_\Clm \approx 1$) until both the length and radius of the channel significantly encroach on the hydration layers. This occurs for a larger radius for bi- and tri-layer graphene as both the first and second hydration layer are more significantly diminished due to the larger pore length. Moreover, $J$ is fairly constant for $r_p$ greater than the second hydration layer radius ($\approx 0.6$ nm) and starts to drop as pore size decreases further. The drop is much sharper below the first hydration layer ($\approx 0.3$ nm) to the extent that we find no current for bi- and tri-layer graphene within the time of our simulations. The error bars are $\pm$ 1 BSE.} 
\end{figure}

\section{``Quantized" ionic current}
Since the ion current density relates to the free energy barrier as $J_\nu  =  J_{\nu0} e^{-\Delta F_\nu / k_B T}$ and the energy barrier is related to the number of waters lost from the solvation shell, the ionic current is expected to have a step-like feature with respect to the pore size, as this determines the extent of dehydration. We see indications of such step-like features in current density, as shown in Fig.~\ref{current}(a). However, the pore sizes themselves are ``discretized''  at this length scale (and not perfectly circular), it is hard to determine if these features are sharp. As we mention in the main text, irregularly shaped nanopores may allow one to examine intermediate pore sizes and determine if these step features are indeed sharp. We leave this for a future study, although it is clear from Fig.~\ref{current}(a) that there is a change in current density when the second (for bi- and tri-layer graphene) and first hydration layers (for all cases) are encroached upon.

\clearpage
\section{Data}

\begin{table}[h!]
\centering
\begin{tabular}{|P{1.5cm} |P{1cm} P{1cm}P{1cm} P{1cm} P{1cm} P{1cm} P{1cm}P{1cm} P{1cm}P{1cm}|}
\hline
     & \multicolumn{10}{c|}{monolayer}                                                        \\ \hline
$r_p$ (nm)    & 0.21  & 0.34  & 0.48  & 0.57  & 0.6   & 0.73  & 0.79  & 0.96  & 1.11  & 1.2   \\ \hline
$I_\Km$ (nA)   & 0.048 & 0.33 & 1.03 & 1.56 & 1.87 & 2.77 & 3.18 & 5.03 & 6.6 & 7.8  \\ \hline
$I_\Clm$ (nA)  & 0.006 & 0.28 & 1.03 & 1.59 & 1.79 & 2.84 & 3.29 & 4.94 & 6.7 & 7.7 \\ \hline
$I_\Km/I_\Clm$    & 8  & 1.2  & 1.0  & 1.0  & 1.0  & 1.0  & 1.0  & 1.0  & 1.0  & 1.0  \\ \hline
     & \multicolumn{10}{c|}{bilayer}                                                          \\ \hline
$r_p$ (nm)    & 0.16  & 0.34  & 0.48  & 0.57  & 0.6   & 0.73  & 0.79  & 0.97  & 1.11  & 1.2   \\ \hline
$I_\Km$ (nA)   & 0     & 0.10 & 0.61 & 0.98 & 1.19 & 1.93 & 2.30 & 3.88 & 5.4 & 6.3 \\ \hline
$I_\Clm$ (nA)  & 0     & 0.04 & 0.60 & 0.99 & 1.23 & 2.09 & 2.55 & 4.06 & 5.5 & 6.3  \\ \hline
$I_\Km/I_\Clm$    & -     & 2.2  & 1.0  & 1.0  & 1.0  & 1.0  & 1.0  & 1.0  & 1.0  & 1.0  \\ \hline
     & \multicolumn{10}{c|}{trilayer}                                                         \\ \hline
$r_p$ (nm)    & 0.14  & 0.34  & 0.48  & 0.57  & 0.6   & 0.73  & 0.79  & 0.97  & 1.11  & 1.2    \\ \hline
$I_\Km$ (nA)   & 0     & 0.059 & 0.45 & 0.82 & 0.95 & 1.59 & 2.05 & 3.28 & 4.8 & 5.4  \\ \hline
$I_\Clm$ (nA)  & 0     & 0.005 & 0.43 & 0.68 & 0.98 & 1.70 & 2.06 & 3.48 & 4.7 & 5.7  \\ \hline
$I_\Km/I_\Clm$    & -     & 11 & 1.0  & 1.2  & 1.0  & 1.0  & 1.0  & 1.0  & 1.0  & 1.0  \\ \hline
\end{tabular}
\caption{\label{currentTable}  K$^+$ and Cl$^-$ currents and their ratio in pores of various radii in mono-, bi- and tri-layer graphene. We measure the currents by counting the ions that cross through the pore. There were no ion crossing events for the smallest pore in bi- and tri-layer graphene. The error in current is $\approx 20$~\% for $r_p=0.21$ nm and $\approx 10$~\% for $r_p=0.34$ nm and $\approx 2$~\% for larger pores.. The error in selectivity is shown in Fig.~\ref{current}.}
\end{table}

\begin{table}[h!]
\centering
\begin{tabular}{|P{2.05cm} |P{1.22cm}| P{1.22cm}|P{1.22cm} |P{1.22cm} |P{1.22cm}| P{1.22cm} |P{1.22cm}|P{1.22cm}|}
\hline
\multirow{3}{*}{}  & \multicolumn{4}{c|}{ K$^+$ } & \multicolumn{4}{c|}{Cl$^-$} \\ \cline{2-9}
& \multicolumn{2}{c|}{ $r_p=0.21$ nm} & \multicolumn{2}{c|}{ $r_p=0.34$ nm} & \multicolumn{2}{c|}{ $r_p=0.21$ nm} & \multicolumn{2}{c|}{ $r_p=0.34$ nm}  \\ \cline{2-9}
 & $n_1$ & $n_2$  & $n_1$ & $n_2$             & $n_1$ & $n_2$           & $n_1$ & $n_2$   \\ \hline
monolayer & 4.7 (3.7) & 13.1 (13.1)    & 7.6 (6.4) & 15.9 (15.6)       & 5.1 (3.8)   & 15.4 (16.6)               & 7.7 (6.6) & 17.8 (15.8)  \\ \hline
bilayer & 3.0 (2.8)  & 5.1 (7.0)    & 7.3 (6.4)   & 9.7 (7.9)         & 2.4 (2.4) & 6.5 (7.0)              & 7.4 (6.6)& 11.1 (7.9) \\ \hline
trilayer & 4.0 (2.8) & 2.1 (2.0)   & 7.3 (6.4) & 8.2 (6.4)           & 4.0 (2.4) & 2.2 (1.7)              & 7.7 (6.6) & 8.8 (5.0) \\ \hline
bulk  & 6.8 & 23.0 & 6.8 & 23.0       & 7.4&  26.3 & 7.4 & 26.3 \\ \hline
\end{tabular}
\caption{\label{WaterTable} The average number of water molecules, $\avg{n}$, in the first and second hydration layer for K$^+$ and Cl$^-$ ions fixed at the center of the two smallest pores and in bulk. The error in $\avg{n}$ is $\approx \pm0.01$ in each case. The estimated water loss considering only the geometric confinement is shown in parentheses. For the geometric estimate, mono-, bi-, and tri graphene is approximated as a cylindrical hole of thickness 0.3, 0.6 and 0.9 nm, respectively.}
\end{table}

\begin{table}[h]
\centering
\begin{tabular}{|P{3.75cm} |P{2.75cm} |P{2.75cm}|P{2.75cm}  |P{2.75cm}| P{2.75cm}}
\hline
\multirow{2}{*}{}  & \multicolumn{2}{c|}{ K$^+$ } & \multicolumn{2}{c|}{Cl$^-$} \\ \cline{2-5}
& $r_p=0.21$ nm &  $r_p=0.34$ nm &  $r_p=0.21$ nm &  $r_p=0.34$ nm  \\ \hline
monolayer & 2.0   & 1.3               & -1.6               & -1.5  \\ \hline
bilayer & 2.2     & 1.4               & -1.9      &-1.5  \\ \hline
trilayer & 2.1    & 1.3               & -1.8               &-1.4  \\ \hline
bulk  & 1.4 & 1.4 & -1.4       & -1.4 \\ \hline
\end{tabular}
\caption{\label{WaterDipoleTable}Average dipole orientation (in Debye) along the radial direction $\avg{p_r}$ in the first hydration layer of $\K$ and $\Cl$ ions fixed in the center of the two smallest pores and in bulk. The total dipole moment of individual water molecule in our model is 2.35 D, thus water molecule in the $r_p=0.21$ nm pore in bi- and tri-layer graphene are almost perfectly oriented along radial direction. The error in $\avg{p_r}$ is $\approx \pm0.01$ in each case. Overall ion concentration is maintained at 1 mol/L in each case. When more water is excluded, especially from the first hydration layer, the remaining water more strongly orients its dipole to energetically compensate for the water loss.}
\end{table}

\clearpage

%